\documentclass[reqno]{mfatshort}
\usepackage{amssymb, latexsym, euscript}
\usepackage[cp1251]{inputenc}
\usepackage[english]{babel}
\usepackage{tikz}
\usetikzlibrary{arrows,decorations.pathmorphing,backgrounds,positioning,fit}

\newtheorem{theorem}{Theorem}[section]
\newtheorem{proposition}[theorem]{Proposition}

\newtheorem{lemma}[theorem]{Lemma}
\newtheorem{definition}[theorem]{Definition}
\newtheorem{corollary}[theorem]{Corollary}
\theoremstyle{definition}

\newtheorem{remark}[theorem]{Remark}

\numberwithin{equation}{section}

\begin{document}

\title{Schr\"{o}dinger Operators
with Non-Symmetric Zero-Range Potentials}

\author[A.~Grod]{A.~Grod}
\address{Institute of Mathematics of the National
Academy of Sciences of Ukraine, Kiev, Ukraine} \email{andriy.grod@yandex.ua}

\author[S.~Kuzhel]{S.~Kuzhel}
\address{AGH University of Science and Technology, Krak\'{o}w, Poland} \email{kuzhel@mat.agh.edu.pl}

\date{DD/MM/2013}
\dedicatory{Dedicated to Professor V.D. Koshmanenko on the occasion of his seventieth birthday.}

\begin{abstract}
Non-self-adjoint Schr\"{o}dinger operators $A_{\mathbf{T}}$ which correspond to non-symmetric zero-range
potentials are investigated. For a given $A_{\mathbf{T}}$, the description of non-real eigenvalues, spectral singularities and exceptional
points are obtained; the possibility of interpretation of $A_{\mathbf{T}}$ as a self-adjoint operator in a Krein space
is studied, the problem of similarity of $A_{\mathbf{T}}$ to a self-adjoint operator in a Hilbert space is solved.
\end{abstract}

\maketitle
\section{Introduction} \label{sec1}
An important class of Schr\"{o}dinger operators is formed by operators with singular perturbations. For example,
this class contains Schr\"{o}dinger operators with zero-range potentials or point interactions. These operators effectively simulate short range interactions and belong to the class of exactly solvable models. Numerous works are devoted to the study of singularly perturbed Schr\"{o}dinger operators, in which a series of approaches to the construction and investigation of such operators are developed (see, e.g.,
\cite{ADKK, AKK, AL1, KKQ, KO} and references therein). These studies, in the majority of cases, deal with \emph{symmetric singular perturbations} that lead to \emph{self-adjoint} Schr\"{o}dinger operators.

In the present paper we study \emph{non-self-adjoint} Schr\"{o}dinger operators which correspond to non-symmetric zero-range
potentials.

Our work was inspirited in part by an intensive development of Pseudo-Hermitian ($\mathcal{PT}$-Symmetric) Quantum Mechanics PHQM  (PTQM) during last decades \cite{B6, B4, MO}. The key point of PHQM/PTQM theories is the
employing of non-self-adjoint operators with certain properties of symmetry
for the description of experimentally observable data.  Briefly speaking, in order to
interpret a given non-self-adjoint operator $A$ in a Hilbert space $\mathfrak{H}$ as a
physically meaning Hamiltonian we have to check the reality of its spectrum and to prove
the existence of a new inner product that ensures the (hidden) self-adjointness of $A$.

The paper is devoted to the implementation of this program for various classes ($\mathcal{PT}$-symmetric operators, $\delta$- and $\delta'$- potentials with complex couplings, see definitions in Examples II-IV of Sec. 2) of non-self-adjoint Schr\"{o}dinger operators $A_{\mathbf{T}}$
corresponding to the Schr\"{o}dinger type differential expression (\ref{lesia11}) with singular zero-range potential
$$
a<\delta,\cdot>\delta(x)+b<\delta',\cdot>\delta(x)+c<\delta,\cdot>\delta'(x)+d<\delta',\cdot>\delta'(x),
$$
where the parameters  $a,b,c,d$ are complex numbers. The matrix ${\mathbf{T}}$ is formed by these parameters and
operators $A_{\mathbf{T}}$ are defined by Lemma \ref{be1}.

In Sec. 2, the necessary and sufficient conditions for the existence of non-real eigenvalues, spectral singularities and exceptional
points of $A_{\mathbf{T}}$ are given.

Sec. 3 is devoted to the interpretation of $A_{\mathbf{T}}$ as a self-adjoint
operator in a Krein space. Such kind of self-adjointness cannot
be considered as completely satisfactory in PHQM/PTQM  because it does not guarantee the unitarity of
the dynamics generated by $A_{\mathbf{T}}$. However, possible realization of $A_{\mathbf{T}}$ as  self-adjoint with respect to some
indefinite metrics (indefinite inner product) allows us to apply well-developed tools of the Krein spaces
theory \cite{AZ} to  solving problem of similarity of $A_{\mathbf{T}}$ to a self-adjoint operator in a Hilbert space.
The similarity property means that $A_{\mathbf{T}}$ turns out to be a self-adjoint operator in a Hilbert space with respect to a
suitable chosen inner product.

In Sec. 4, we solve the similarity problem for $A_{\mathbf{T}}$ with the use of a general criterion of similarity \cite{NA} and the Krein spaces methods.

The properties of $A_{\mathbf{T}}$ established in the paper illustrate a typical PHQM/PTQM evolution of spectral properties which can be obtained by
changing entries of ${\mathbf{T}}$: complex eigenvalues $\to$ spectral singularities / exceptional points $\to$  similarity to a self-adjoint operator. For this reason, the operators $A_{\mathbf{T}}$ considered in the work can be used as exactly solvable models of
PHQM/PTQM.

Throughout the paper $\mathcal{D}(A)$, $\mathcal{R}(A)$, and
$\ker{A}$ denote the domain, the range, and the null-space of a
linear operator $A$, respectively, while
$A\upharpoonright{\mathcal{D}}$ stands for the restriction of $A$ to
the set $\mathcal{D}$.  The resolvent set and the spectrum of an operator $A$ are denoted as $\rho(A)$ and $\sigma(A)$, respectively.

\setcounter{equation}{0}
\section{Operator realizations and their simplest properties.}
\label{sec2}
A one-dimensional Schr\"{o}dinger operator corresponding to a
general zero-range potential at the point $x=0$ can be defined by
the heuristic expression
 \begin{equation}\label{lesia11}
 -\frac{d^2}{dx^2}+a<\delta,\cdot>\delta(x)+b<\delta',\cdot>\delta(x)+
 c<\delta,\cdot>\delta'(x)+d<\delta',\cdot>\delta'(x),
 \end{equation}
where $\delta$ and $\delta'$ are, respectively, the Dirac $\delta$-function
and its derivative (with support at $0$) and $a,b,c,d$ are complex
numbers.

The expression (\ref{lesia11}) gives rise to the symmetric operator
\begin{equation}\label{tato1}
A_{\mathrm{sym}}=-\frac{d^2}{dx^2}, \hspace{5mm}
\mathcal{D}(A_{\mathrm{sym}})=\{u(x)\in{W_2^2}(\mathbb{R})\ | \
u(0)=u'(0)=0\}
\end{equation}
acting in $L_2(\mathbb{R})$ and, generally speaking, any proper extension $A$ of
$A_{\mathrm{sym}}$ (i.e., $A_{\mathrm{sym}}\subset{A}\subset{A}_{\mathrm{sym}}^*$ )
can be considered as an operator realization of (\ref{lesia11}) in $L_2(\mathbb{R})$.

In order to specify more exactly which a proper extension
$A$ of $A_{\mathrm{sym}}$ corresponds to (\ref{lesia11})
we will use an approach suggested in \cite{AL1}.
The idea consists in the construction of
some regularization $\mathbb{A}_{\mathbf{r}}$ of (\ref{lesia11}) that
is well defined as an operator from
$\mathcal{D}(A_{\mathrm{sym}}^*)=W_2^2(\mathbb{R}\backslash\{0\})$ to
$W_2^{-2}(\mathbb{R})$. Then, the corresponding operator realization
 of (\ref{lesia11}) in $L_2(\mathbb{R})$ is determined as follows:
\begin{equation}\label{lesia40}
A=\mathbb{A}_\mathbf{r}\upharpoonright_{\mathcal{D}(A)}, \ \ \ \
\mathcal{D}(A)=\{f\in\mathcal{D}(A_{\mathrm{sym}}^*) \ | \
\mathbb{A}_\mathbf{r}f\in{L_2(\mathbb{R})}\}.
\end{equation}

To obtain a regularization of (\ref{lesia11}) it suffices to extend
the distributions $\delta$ and $\delta'$ onto
${W_2^2}(\mathbb{R}\backslash\{0\})$.
The most reasonable way (based on preserving of initial homogeneity
of $\delta$ and $\delta'$ with respect to scaling transformations,
see, for details, \cite{AL1}, \cite{HK}) leads to the following
definition:
$$
 <\delta_{ex}, f>=\frac{f(+0)+f(-0)}{2}, \hspace{3mm} <\delta_{ex}', f>=-\frac{f'(+0)+f'(-0)}{2}
$$
for all $f(x)\in{{W_2^2}(\mathbb{R}\backslash\{0\})}$. In this case, the
regularization of (\ref{lesia11}) onto
${W_2^2}(\mathbb{R}\backslash\{0\})$ takes the form
$$
\mathbb{A}_\mathbf{r}=-\frac{d^2}{dx^2}+a<\delta_{ex},\cdot>\delta(x)+b<\delta_{ex}',\cdot>\delta(x)+
c<\delta_{ex},\cdot>\delta'(x)+d<\delta_{ex}',\cdot>\delta'(x),
$$
where $-{d^2}/{dx^2}$ acts on ${W_2^2}(\mathbb{R}\backslash\{0\})$ in
the distributional sense.

The definition (\ref{lesia40}) is not always easy to use.
Repeating the proof of Theorem 1 in \cite{AK2} we obtain an equivalent description of
operators determined by (\ref{lesia40}).
\begin{lemma}\label{be1}
Let $A$ be determined by (\ref{lesia40}). Then  $A$ coincides with the restriction of
 $A_{\mathrm{sym}}^*=-d^2/dx^2$ onto the domain
 \begin{equation}\label{lesia800}
 \mathcal{D}(A)=\{f(x)\in{{W_2^2}(\mathbb{R}\backslash\{0\})} \  \mid  \
 \mathbf{T}\Gamma_0f=\Gamma_1f \},  \ \ \ \ \  \mathbf{T}=\left(\begin{array}{cc}
 a & b \\
 c & d
 \end{array}\right),
 \end{equation}
where
\begin{equation}\label{sas70}
\Gamma_0f=\frac{1}{2}\left(\begin{array}{c}
 f(+0)+f(-0) \\
-f'(+0)-f'(-0)
\end{array}\right), \hspace{8mm}  \Gamma_1f=\left(\begin{array}{c}
f'(+0)-f'(-0) \\
f(+0)-f(-0)
\end{array}\right).
\end{equation}
\end{lemma}

\begin{remark}
In what follows the notation $A_{\mathbf{T}}$ will be used for operator realizations of (\ref{lesia11})
defined by (\ref{lesia800}) and (\ref{sas70}).
\end{remark}

It is known that the continuous spectrum of an operator $A_{\mathbf{T}}$
coincides with $[0,\infty)$ and the point spectrum of
$A_{\mathbf{T}}$ may appear only in $\mathbb{C}\backslash\mathbb{R}_+$.

Denote
\begin{equation}\label{bbb2}
 \mathbf{{det} \ T}=ad-bc.
\end{equation}
\begin{lemma}\label{l10}
An operator $A_{\mathbf{T}}$ has an eigenvalue $z=\tau^2$ if and only if
the equation
\begin{equation}\label{lesia505}
2d\tau^2+i(\mathbf{{det} \ T}-4)\tau+2a=0
\end{equation}
has a  solution $\tau\in\mathbb{C}_+=\{\tau\in\mathbb{C} \ : \ \textsf{Im}\ \tau>0 \}$.
\end{lemma}

\emph{Proof.}  Let us denote by $\tau$  the square root
of the energy parameter $z=\tau^2$ determined uniquely by
the condition $\textsf{Im}\ \tau>0$ and consider the functions
\begin{equation}\label{sas2}
 h_{1\tau}(x)=\left\{\begin{array}{cc}
 e^{i\tau{x}}, & x>0;  \\
 e^{-i\tau{x}}, & x<0
 \end{array}\right.    \hspace{10mm}
 h_{2\tau}(x)=\left\{\begin{array}{cc}
 -e^{i\tau{x}} , & x>0  \\
 e^{-i\tau{x}}, & x<0
 \end{array}\right.
 \end{equation}
that form a basis of  $\ker(A_{\mathrm{sym}}^*-zI)$, where $z=\tau^2$ runs
$\mathbb{C}\backslash\mathbb{R}_+$.  It is clear that $z$ belongs
to the point spectrum of $A$ if and only if there exists a
function $f\in\ker(A_{\mathrm{sym}}^*-zI)\cap\mathcal{D}(A)$.
Representing $f(x)$ in the form
$$
f(x)=c_1h_{1\tau}(x)+c_2h_{2\tau}(x) \quad c_i\in\mathbb{C}
$$
and substituting this expression into (\ref{lesia800}) we arrive at the conclusion that $z$ is an
eigenvalue of $A$ if and only if the system of equations
$$
\begin{array}{c}
 (a-2i\tau)c_{1}+ib{\tau}c_{2}=0 \\
 {c}c_{1}+(id\tau+2)c_{2}=0
\end{array}
$$
has a nontrivial solution $c_1, c_2$. This is possible
if the determinant of the coefficient matrix of the system
is equal to zero, i.e.,
$2d\tau^2+i(ad-bc-4)\tau+2a=0.$ Rewriting the obtained equation in the form
(\ref{lesia505}) we complete the proof.
\rule{2mm}{2mm}
\begin{definition}\label{yahoo}
Let $A_{\mathbf{T}}$ be defined by (\ref{lesia800}), (\ref{sas70}) and let the spectrum of $A_{\mathbf{T}}$
do not coincide with $\mathbb{C}.$
We will say that the operator $A_{\mathbf{T}}$ has:
\begin{itemize}
  \item \emph{a nonzero spectral singularity} $z=\tau^2$ if
the equation (\ref{lesia505}) has a  solution $\tau\in\mathbb{R}\setminus\{0\}$;
\item \emph{a spectral singularity at point}
$z=0$ if (\ref{lesia505}) has a solution $\tau=0$ with multiplicity $2$;
\item \emph{a spectral singularity at point}
$z=\infty$ if there are no solutions of (\ref{lesia505}) in $\mathbb{C}$.
\end{itemize}
 The non-self-adjoint operator $A_{\mathbf{T}}$ has \emph{an exceptional point} $z=\tau^2$ if
the equation (\ref{lesia505}) has a  solution $\tau\in\mathbb{C}_{+}$ with
multiplicity $2$.
\end{definition}

A spectral singularity (an exception point) $z$  lies on the continuous spectrum (on the point spectrum) of $A_{\mathbf{T}}$ and
it is a serious defect that rules out the operator as a viable candidate for a physical observable \cite{GR, MO1}.

{\bf Example I.} \emph{Symmetric potential.}

The singular potential
\begin{equation}\label{bbb4}
V=a<\delta,\cdot>\delta(x)+b<\delta',\cdot>\delta(x)+
 c<\delta,\cdot>\delta'(x)+d<\delta',\cdot>\delta'(x)
\end{equation} in (\ref{lesia11})
 is symmetric (i.e., $V^*=V$) if and only if
 \begin{equation}\label{bbb3}
 a, \ d\in{\mathbb{R}}, \qquad  c=\overline{b}.
\end{equation}
The corresponding operators $A_{\mathbf{T}}$  turn out to be  self-adjoint operators in $L_2(\mathbb{R})$
with respect to the initial inner product
\begin{equation}\label{bbb8}
(f,g)=\int_{\mathbb{R}}f(x)\overline{g(x)}dx.
\end{equation}

\begin{lemma}
The spectrum $\sigma(A_{\mathbf{T}})$ is real and it contains the continuous part $[0,\infty)$ and possibly, negative eigenvalues.
There are no spectral singularities and exceptional points.
\end{lemma}
\emph{Proof.}
An operator $A_{\mathbf{T}}$ is a finite dimensional extension of the symmetric operator
$A_{\mathrm{sym}}$ determined by (\ref{tato1}).  This means that the continuous spectrum of $A_{\mathbf{T}}$ coincides  with
$[0,\infty)$.

 Let $d\not=0$. Then the solutions $\tau_{1,2}$ of (\ref{lesia505}) have the form
\begin{equation}\label{bbb1}
\tau_{1,2}=i\frac{4-\mathbf{{det} \ T}\pm\sqrt{D}}{4d},
\end{equation}
where $\mathbf{{det} \ T}$ and $D=(4-\mathbf{{det} \ T})^2+16ad$
are real numbers.

Taking (\ref{bbb2}) into account we rewrite
\begin{equation}\label{bbb17}
D=(4-ad+bc)^2+16ad=(4+ad-bc)^2+16bc=(4+\mathbf{{det} \ T})^2+16bc.
\end{equation}
Moreover, in view of (\ref{bbb3}), $bc=|b|^2$. Therefore $D=(4+\mathbf{{det} \ T})^2+16|b|^2\geq{0}$.
This means that the solutions $\tau_{1,2}$ determined by (\ref{bbb1}) \emph{always belong to} $i\mathbb{R}$.

Similarly, if $d=0$, equation (\ref{lesia505}) is reduced to
$-i(|b|^2+4)\tau+2a=0$. The solution $\tau_1=-2ai/(|b|^2+4)$  \emph{belongs to} $i\mathbb{R}$.

The two cases above and Lemma \ref{l10} show that $A_{\mathbf{T}}$ may have negative eigenvalues $z=\tau^2$  and
there are no spectral singularities and exceptional points of $A_{\mathbf{T}}$. \rule{2mm}{2mm}

{\bf Example II.} \emph{$\mathcal{PT}$-symmetric potential.}

Denote by ${\mathcal P}$ and ${\mathcal T}$ the operators of space parity and complex conjugation, respectively:
\begin{equation}\label{bbb11}
{\mathcal P}f(x)=f(-x), \qquad {\mathcal T}f(x)=\overline{f(x)}.
\end{equation}

The potential $V$ is called $\mathcal{PT}$-symmetric if $\mathcal{PT}V=V\mathcal{PT}$.
Extending ${\mathcal P}$ onto $W_2^{-2}(\mathbb{R})$, one gets
${\mathcal P}\delta=\delta$ and ${\mathcal P}\delta'=-\delta'$.
These relations and (\ref{bbb4}) imply
that $V$ is $\mathcal{PT}$-symmetric if and only if\footnote{the cases of symmetric and  $\mathcal{PT}$-symmetric potentials differs by conditions imposed on parameters $b, c$.}
\begin{equation}\label{bbb35}
a, \ d\in\mathbb{R}, \qquad \ b, \ c \in{i\mathbb{R}}.
\end{equation}

The corresponding operators $A_{\mathbf{T}}$  turn out to be $\mathcal{PT}$-symmetric operators, i.e., the relation:
\begin{equation}\label{bbb5}
\mathcal{P}\mathcal{T}A_{\mathbf{T}}=A_{\mathbf{T}}\mathcal{P}\mathcal{T}
\end{equation}
holds on the domain $\mathcal{D}(A_{\mathbf{T}})$.

\begin{remark}\label{neww89}
In what follows we will often use
operator identities
\begin{equation}\label{neww11}
XA=BX,
\end{equation}
where $A$ and $B$ are (possible) unbounded operators in $L_2(\mathbb{R})$ and  $X$ is a bounded operator in $L_2(\mathbb{R})$.
In that case, we \emph{always assume that (\ref{neww11}) holds on $\mathcal{D}(A)$}. This means that
$X : \mathcal{D}(A)\to\mathcal{D}(B)$ and the identity $XAu=BXu$ holds for all $u\in\mathcal{D}(A)$.
If $A$ is bounded, then (\ref{neww11}) should hold on the whole $L_2(\mathbb{R})$.
In particular, relation (\ref{bbb35}) means that the operator $\mathcal{P}\mathcal{T}$ maps $\mathcal{D}(A_{\mathbf{T}})$ onto $\mathcal{D}(A_{\mathbf{T}})$
and $\mathcal{P}\mathcal{T}A_{\mathbf{T}}f=A_{\mathbf{T}}\mathcal{P}\mathcal{T}f$ for all $f\in\mathcal{D}(A_{\mathbf{T}})$.
\end{remark}

Comparing the condition of self-adjointness (\ref{bbb3}) and the condition of $\mathcal{P}\mathcal{T}$-symmetry (\ref{bbb5})
we obtain that  $\mathcal{P}\mathcal{T}$-symmetric operators $A_{\mathbf{T}}$
are not self-adjoint with respect to the initial inner product (\ref{bbb8}) except the case $b=-c\in{i\mathbb{R}}$.
Therefore, $\mathcal{P}\mathcal{T}$-symmetric operators $\mathcal{D}(A_{\mathbf{T}})$ may have
non-real eigenvalues. In particular, it may happen that the set of complex eigenvalues of $A_{\mathbf{T}}$
coincide with $\mathbb{C}\setminus\mathbb{R}_+$.

\begin{lemma}\label{kkk1}
1. A $\mathcal{P}\mathcal{T}$-symmetric operator $A_{\mathbf{T}}$ has \emph{non-real eigenvalues} if and only if
one of the following conditions are satisfied:
$$
  (i) \quad D=(4-\mathbf{{det} \ T})^2+16ad<0, \qquad  (4-\mathbf{{det} \ T})d>0;
$$
$$
  (ii) \quad \mathbf{{det} \ T}=4, \qquad a=d=0.
$$
Condition (i) corresponds to the case where $A_{\mathbf{T}}$ has two non-real
eigenvalues, which are conjugate to each other.
Condition (ii) describes the situation where any point
$z\in\mathbb{C}\backslash\mathbb{R}_+$ is an eigenvalue of $A_{\mathbf{T}}$. In that case the spectrum of $A_{\mathbf{T}}$
coincides with $\mathbb{C}$;

2. A $\mathcal{P}\mathcal{T}$-symmetric operator $A_{\mathbf{T}}$ has:
\begin{itemize}
\item \emph{nonzero spectral singularity} if and only if
$$
(iii) \quad D<0, \quad (4-\mathbf{{det} \ T})d=0
$$
In that case, the positive number $z=-\frac{a}{d}$ is the spectral singularity of $A_{\mathbf{T}}$;

\item \emph{spectral singularity at point $z=0$} if and only if
$$
(iv) \quad D=0, \quad (4-\mathbf{{det} \ T})d=0, \quad d\not=0, \quad a=0;
$$

\item \emph{spectral singularity at point $z=\infty$} if and only if
$$
(v) \quad D=0, \quad (4-\mathbf{{det} \ T})d=0, \quad d=0, \quad a\not=0;
$$

\item \emph{exceptional point} if and only if
$$
(vi) \quad  D=0, \quad (4-\mathbf{{det} \ T})d>0.
$$
In that case, the negative number $z=\frac{a}{d}$ is the exceptional point of $A_{\mathbf{T}}$.
\end{itemize}
\end{lemma}

\emph{Proof.} Let $A_{\mathbf{T}}$ be a $\mathcal{P}\mathcal{T}$-symmetric operator. Then,
the values of $\mathbf{{det} \ T}$ and $D$ are real (it follows from (\ref{bbb35})).

Using Lemma \ref{l10} and (\ref{bbb1}), we arrive at the conclusion that
condition (i) is necessary and sufficient for
the existence of two non-real eigenvalues $z_{1,2}=\tau^2_{1,2}$ of
$A_{\mathbf{T}}$, which are conjugate to each other.

The requirement that any point
$z=\tau^2\in\mathbb{C}\backslash\mathbb{R}_+$ is an eigenvalue of $A_{\mathbf{T}}$ is equivalent
to the condition that any $\tau\in\mathbb{C}_+$ is a solution of
(\ref{lesia505}). This is possible only in the case where
the left-hand side of (\ref{lesia505}) vanishes. The latter is equivalent to the
 condition (ii).

 It follows from (\ref{bbb1}) that a nonzero spectral singularity exists in the case where $D=(4-\mathbf{{det} \ T})^2+16ad<0$ and
$\mathbf{{det} \ T}=4$ that is equivalent to (iii). In that case, $\tau_{1,2}=\pm\sqrt{-\frac{a}{d}}$ and
$z=\tau_{1,2}^2=-\frac{a}{d}$.

The descriptions (iv) and (v) of spectral singularities at $0$ and at $\infty$ are obvious due to (\ref{lesia505}).

By Definition \ref{yahoo} a point $z=\tau^2$ ($\tau\in\mathbb{C}_+$) is an exceptional point of
$A_{\mathbf{T}}$ if its multiplicity is $2$. Hence, $d\not=0$ in (\ref{lesia505}) and $\tau$ is determined by (\ref{bbb1})
with $D=0$. Furthermore, the condition $\tau\in\mathbb{C}_+$
implies $(4-\mathbf{{det} \ T})d>0$. Thus we show that condition (vi) corresponds to exceptional points.
In that case
$$
z=\tau^2=-\frac{(4-\mathbf{{det} \ T})^2}{16d^2}=\frac{16ad}{16d^2}=\frac{a}{d}<0
$$
(since $D=0$, $(4-\mathbf{{det} \ T})\not=0$ and hence, $ad<0$).
 Lemma \ref{kkk1} is proved.
\rule{2mm}{2mm}

{\bf Example III.} \emph{$\delta$-potential with a complex coupling \cite{MO1}.} \\
Let $a\in\mathbb{C}$ and $b=c=d=0$. Then (\ref{lesia11}) takes the form
$$
 -\frac{d^2}{dx^2}+a<\delta,\cdot>\delta(x), \qquad a\in\mathbb{C}
$$
and (\ref{lesia800}) gives rise to operators $A_{\mathbf{T}}\equiv{A_a}=\displaystyle{-\frac{d^2}{dx^2}}$ with domains of definition
$$
 \mathcal{D}(A_a)=\left\{f(x)\in{{W_2^2}(\mathbb{R}\backslash\{0\})} \  \left|\right.  \begin{array}{l}
 f(0+)=f(0-) \ (\equiv{f(0)}) \vspace{2mm} \\
 f'(0+)-f'(0-)=af(0)\end{array} \right\}.
 $$
By virtue of Lemma \ref{l10} and Definition \ref{yahoo} we conclude:
\begin{itemize}
  \item if $\textsf{Re}\ a<0$, then $A_a$ has a unique eigenvalue $z=-a^2/4$, which is real $\iff$ $\textsf{Im}\ a=0$;
  \item if $\textsf{Re}\ a\geq0$, then the spectrum of $A_a$ is real, continuous and it coincides with $[0,\infty)$;
\item if $a\in{i}\mathbb{R}\setminus\{0\}$, then $A_a$ has spectral singularity $z=\frac{|a|^2}{4}$;
\item there are no exceptional points of $A_a$.
\end{itemize}

{\bf Example IV.} \emph{$\delta'$-potential with a complex coupling.} \\
Let $d\in\mathbb{C}$ and $a=b=c=0$. Then (\ref{lesia11}) takes the form
$$
 -\frac{d^2}{dx^2}+d<\delta',\cdot>\delta'(x), \qquad d\in\mathbb{C}
$$
and (\ref{lesia800}) gives rise to operators $A_{\mathbf{T}}\equiv{A_d}=\displaystyle{-\frac{d^2}{dx^2}}$ with domains of definition
$$
 \mathcal{D}(A_d)=\left\{f(x)\in{{W_2^2}(\mathbb{R}\backslash\{0\})} \  \left|\right. \begin{array}{l}
 f'(0+)=f'(0-) \ (\equiv{f'(0)}) \vspace{2mm} \\
 f(0+)-f(0-)=-df'(0)
 \end{array} \right\}
 $$
By virtue of Lemma \ref{l10} and Definition \ref{yahoo}:
\begin{itemize}
  \item if $\textsf{Re}\ d>0$, then $A_d$ has a unique eigenvalue $z=-4/d^2$, which is real $\iff$ $\textsf{Im}\ d=0$;
  \item if $\textsf{Re}\ d\leq0$, then the spectrum of $A_d$ is real, continuous and it coincides with $[0,\infty)$;
\item if $d\in{i}\mathbb{R}\setminus\{0\}$, then $A_d$ has spectral singularity $z=\frac{4}{|d|^2}$;
\item there are no exceptional points of $A_d$.
\end{itemize}

\section{Interpretation as Self-Adjoint Operators in Krein Spaces.}

Let $\mathfrak{H}$ be a Hilbert space with inner product
$(\cdot,\cdot)$ and with fundamental symmetry $J$ (i.e., $J=J^*$ and
$J^2=I$). The space $\mathfrak{H}$ endowed with the indefinite inner product
(indefinite metric) \ $[f,g]_J:=(J{f}, g), \ \forall{f,g}\in\mathfrak{H}$ is
called  a Krein space  $(\mathfrak{H}, [\cdot,\cdot]_J)$.

The difference between the initial inner product $(\cdot,\cdot)$ and indefinite metric
$[\cdot,\cdot]_J$ consists in the fact that, except the cases $J=\pm{I}$, the sign of the sesqulinear form $[f,f]_J$
is not determined (i.e, it is possible $[f,f]_J<0$, $[f,f]_J=0$, or $[f,f]_J>0$ for various $f\not=0$).
The Hilbert space $\mathfrak{H}$
can be considered as a particular case of the Krein space $(\mathfrak{H}, [\cdot,\cdot]_J)$ with $J=I$.

A linear densely defined\footnote{with respect to the initial inner product $(\cdot,\cdot)$} operator $A$ acting in $\mathfrak{H}$ is called
self-adjoint in the Krein space  $(\mathfrak{H}, [\cdot,\cdot]_J)$ if $A$ is
self-adjoint with respect to the indefinite metric $[\cdot,\cdot]_J$. This condition is equivalent to the relation
\begin{equation}\label{bbb23}
A^*={J}A{J}.
\end{equation}

The spectrum of a self-adjoint operator in Krein space is symmetric
with respect to the real axis. For an additional information about Krein spaces and operators acting therein
we refer to \cite{AZ}.

 We recall \cite{MO, M1} that a linear densely defined operator
$A$ acting in a Hilbert space ${\mathfrak H}$  is said to be
{\it pseudo-Hermitian} if there exists a bounded and boundedly invertible self-adjoint operator
$\eta : {\mathfrak H} \to {\mathfrak H}$  such that
 \begin{equation}\label{e1}
             {A}^*=\eta{A}\eta^{-1}.
 \end{equation}

Relation (\ref{e1}) means that $A$ is self-adjoint with respect to the pseudo-metric $[\cdot,\cdot]_{\eta}=(\eta\cdot,\cdot)$.

It follows from (\ref{bbb23}) and (\ref{e1}) that
self-adjoint operators in Krein spaces are pseudo-Hermitian. The inverse implication is also true.
Indeed, let $A$ be pseudo-Hermitian. Then (\ref{e1}) holds for some $\eta$. Denote
\begin{equation}\label{rrr2}
|\eta|=\sqrt{\eta^2}, \qquad J=\eta|\eta|^{-1}
\end{equation}
and consider the Hilbert space $(\mathfrak{H}, (\cdot,\cdot)_{|\eta|})$ endowed with new (equivalent to $(\cdot,\cdot)$) inner product
$(\cdot,\cdot)_{|\eta|}=(|\eta|\cdot,\cdot)$. Then, the pseudo-metric $[\cdot,\cdot]_{\eta}$ coincides with the indefinite metric
$[\cdot,\cdot]_{J{|\eta|}}=(J\cdot,\cdot)_{|\eta|}$ constructed with the use of fundamental symmetry $J=\eta|\eta|^{-1}$ and
new inner product $(\cdot,\cdot)_{|\eta|}$, i.e.,
$$
[\cdot,\cdot]_{\eta}=(\eta\cdot,\cdot)=(J|\eta|\cdot,\cdot)=(J\cdot,\cdot)_{|\eta|}=[\cdot,\cdot]_{J{|\eta|}}.
$$ This means that $A$ turns out to be a self-adjoint operator in the Krein space
$(\mathfrak{H}, [\cdot,\cdot]_{J{|\eta|}})$.

{\bf Example II contd.} It is known \cite{PK} that an arbitrary ${\mathcal P}{\mathcal T}$-symmetric operator $A_{\mathbf{T}}$
can be interpreted as self-adjoint one in a suitable chosen Krein space $(L_2(\mathbb{R}), [\cdot,\cdot]_J)$.
Using \cite{PK} we can specify the relevant indefinite metrics $[\cdot,\cdot]_J$. Denote
\begin{equation}\label{bbb34}
{\mathcal R}f(x)=\textsf{sign}(x)f(x), \qquad f\in{L_2({\mathbb R})}.
\end{equation}
The operator ${\mathcal R}$ is a fundamental symmetry which anti-commutes with $\mathcal{P}$: $\mathcal{P}\mathcal{R}=-\mathcal{R}\mathcal{P}$. It is easy to check that the operator $i\mathcal{PR}$ is also a fundamental symmetry and, moreover, any operator
\begin{equation}\label{bebebe}
J_{\vec{\alpha}}=\alpha_1\mathcal{P}+\alpha_2\mathcal{R}+\alpha_3i\mathcal{PR}, \qquad \alpha_j\in\mathbb{R}, \quad \alpha_1^2+\alpha_2^2+\alpha_3^2=1
\end{equation}
turns out to be a fundamental symmetry in $L_2(\mathbb{R})$.

We consider also a subset of the set of fundamental symmetries
 $J_{\vec{\alpha}}$ by imposing an additional condition of $\mathcal{PT}$-symmetry:
 $\mathcal{PT}J_{\vec{\alpha}}=J_{\vec{\alpha}}\mathcal{PT}$.

 The operator
 $J_{\vec{\alpha}}$ is  $\mathcal{PT}$-symmetric if and only if $\alpha_2=0$.
In that case the latter relation in (\ref{bebebe}) takes the form $\alpha_1^2+\alpha_3^2=1$ and we may set
$\alpha_1=\cos\phi$ and $\alpha_3=\sin\phi$. Then
$$
J_{\vec{\alpha}}=(\cos\phi)\mathcal{P}+i(\sin\phi)\mathcal{PR}=\mathcal{P}(\cos\phi+i(\sin\phi)\mathcal{R})=\mathcal{P}e^{i\phi\mathcal{R}}.
$$
Thus, fundamental symmetries $J_{\vec{\alpha}}$ \emph{with the additional property of $\mathcal{PT}$-symmetry coincide
with fundamental symmetries} $\mathcal{P}_\phi=\mathcal{P}e^{i\phi\mathcal{R}}, \ \phi\in[0,2\pi)$.

Consider the following collection of indefinite metrics on $L_2(\mathbb{R})$:
$$
[\cdot,\cdot]_{J_{\vec{\alpha}}}=(J_{\vec{\alpha}}\cdot, \cdot),  \qquad [\cdot,\cdot]_{\mathcal{P}_\phi}=(\mathcal{P}_\phi\cdot, \cdot).
$$
\begin{proposition}[\cite{PK}]\label{hhh1}
 Every ${\mathcal{PT}}$-symmetric operator  $A_\mathbf{T}$ can be interpreted as a self-adjoint operator
 in the Krein space $(L_2(\mathbb{R}), [\cdot,\cdot]_{\mathcal{P}_\phi})$, where the parameter
  $\phi$ is determined by the relation
\begin{equation}\label{rere1}
2(b-c)\cos\phi=i(4+\det\mathbf{T})\sin\phi.
\end{equation}
\end{proposition}

Proposition \ref{hhh1} shows that the collection of Krein spaces $(L_2(\mathbb{R}), [\cdot,\cdot]_{\mathcal{P}_\phi})$ generated by  $\mathcal{PT}$-symmetric fundamental symmetries
$\mathcal{P}_\phi$ \emph{is sufficient for the interpretation of $A_\mathbf{T}$ as a self-adjoint operator}.
The possible interpretation of some $A_\mathbf{T}$ as a self-adjoint operator in a Krein space
$(L_2(\mathbb{R}), [\cdot,\cdot]_{J_{\vec{\alpha}}})$, where $J_{\vec{\alpha}}$ \emph{is not ${\mathcal{PT}}$-symmetric}
(i.e. $J_{\vec{\alpha}}\not=\mathcal{P}_\phi$) immediately leads to specific spectral properties of $A_\mathbf{T}$.
\begin{proposition}[\cite{PK}]\label{hhh2}
Let $A_\mathbf{T}$ be a non-self-adjoint $\mathcal{PT}$-symmetric operator. Then
\begin{itemize}
  \item if $A_\mathbf{T}$ admits an interpretation as a self-adjoint operator in a Krein space $(L_2(\mathbb{R}), [\cdot,\cdot]_{J_{\vec{\alpha}}})$, where $J_{\vec{\alpha}}$ is not ${\mathcal{PT}}$-symmetric, then $\sigma(A_\mathbf{T})=\mathbb{C}$;
  \item if $A_\mathbf{T}$ admits an interpretation as a self-adjoint operator in two different Krein spaces
  $(L_2(\mathbb{R}), [\cdot,\cdot]_{\mathcal{P}_{\phi_1}})$ and $(L_2(\mathbb{R}), [\cdot,\cdot]_{\mathcal{P}_{\phi_2}})$, where
  ${\mathcal{P}_{\phi_1}}$ and ${\mathcal{P}_{\phi_2}}$ are \emph{linearly independent}, then the spectrum of $A_\mathbf{T}$ contains a pair
  of complex conjugated eigenvalues;
  \item if $A_\mathbf{T}$ has a real spectrum, then $A_\mathbf{T}$ has interpretation as self-adjoint operator for the \emph{unique choice}\footnote{up to linearly dependent fundamental symmetries $\mathcal{P}_{\phi}$}
  of the Krein space $(L_2(\mathbb{R}), [\cdot,\cdot]_{\mathcal{P}_{\phi}})$.
\end{itemize}
\end{proposition}

{\bf Examples III. VI contd.}
It follows from (\ref{bbb3}) and (\ref{bbb35}) that
operators $A_a$ ($A_d$) with real $a$ (real $d$) are self-adjoint in the initial Hilbert space $L_2(\mathbb{R})$.
Moreover, taking \cite{AK2} into account, we decide that these operators are self-adjoint in the Krein space $(L_2(\mathbb{R}), [\cdot,\cdot]_{\mathcal P})$, where
\begin{equation}\label{neww78}
[f,g]_{\mathcal P}=({\mathcal P}f,g)=\int_{\mathbb{R}}f(-x)\overline{g(x)}dx.
\end{equation}

If $a$ is non-real and $\textsf{Re}\ a<0$  (if $d$ is non-real and $\textsf{Re}\ d>0$), then
$A_a$ ($A_d$) \emph{cannot be interpreted as pseudo-Hermitian operator}, or that is equivalent,
\emph{cannot be interpreted as a self-adjoint operator in a Krein space}. Indeed, if we assume that such an interpretation is possible, then
the spectrum of $A_a$ (of $A_d$) must be symmetric with respect to the real axis that contradicts to the fact that
the spectrum of $A_a$ (of $A_d$) contains \emph{a unique complex eigenvalue} $z=-a^2/4$ \ ($d=-4/d^2$).

If $a\in{i\mathbb{R}}\setminus\{0\}$, ($d\in{i\mathbb{R}}\setminus\{0\}$) the operator $A_a$ ($A_d$)
has a spectral singularity. Hence, $A_a$ ($A_d$) cannot be interpreted as a self-adjoint in a Hilbert space (see Theorem \ref{bbb26}).
The problem of interpretation of $A_a$ ($A_d$) as a self-adjoint operator in a Krein space is still open.

If $\textsf{Re}\ a>0$ ($\textsf{Re}\ d<0$), then the operator $A_a$ ($A_d$) turns out to be self-adjoint in $L_2(\mathbb{R})$
for a certain choice of inner product equivalent to the initial one $(\cdot,\cdot)$ (it follows from  Corollary \ref{bbb27}).

\section{Similarity to Self-Adjoint Operators}
 An operator $A$ acting in a Hilbert space $\mathfrak{H}$ is called \emph{similar} to a
 self-adjoint operator $H$  if there exists a bounded and boundedly invertible operator
 $Z$ such that
 \begin{equation}\label{bbbrrr1}
 A=Z^{-1}HZ.
\end{equation}
The similarity of $A$ to a self-adjoint operator means that $A$ \emph{turns out to be
self-adjoint for a certain choice of inner product of $\mathfrak{H}$, which is equivalent
to the initial inner product $(\cdot,\cdot)$}. Indeed, let (\ref{bbbrrr1}) hold.
By analogy with (\ref{rrr2}) we denote
\begin{equation}\label{rrr3}
|Z|=\sqrt{Z^*Z}, \qquad U=Z|Z|^{-1}
\end{equation}
and rewrite (\ref{bbbrrr1}) as follows:
$$
H=ZAZ^{-1}=U|Z|A|Z|^{-1}U^{-1}=UKU^{-1}, \qquad K=|Z|A|Z|^{-1}.
$$
The operator $U$ is unitary but, in general, $U$ is not self-adjoint.\footnote{this is a difference with the operator $J$ in (\ref{rrr2}).}
Taking into account that $H$ is self-adjoint, we obtain
$$
H^*=(UKU^{-1})^*=UK^*U^{-1}=H=UKU^{-1}.
$$
Therefore, $K^*=K$. Then
$$
(|Z|A|Z|^{-1})^*=|Z|^{-1}A^*|Z|=|Z|A|Z|^{-1}
$$ or, that is equivalent
$A^*|Z|^2=|Z|^2A$. The obtained relation allows us to prove the self-adjointness of $A$
in the Hilbert space $(\mathfrak{H}, (\cdot,\cdot)_{|Z|^2})$ endowed with new (equivalent to $(\cdot,\cdot)$) inner product
$(\cdot,\cdot)_{|Z|^2}=(|Z|^2\cdot,\cdot)$. Indeed,
$$
(Af,g)_{|Z|^2}=(|Z|^2Af,g)=(f, A^*|Z|^2g)=(f,|Z|^2Ag)=(f,Ag)_{|Z|^2}, \quad f,g\in\mathcal{D}(A).
$$
Thus $A$ is self-adjoint in the Hilbert space $(\mathfrak{H}, (\cdot,\cdot)_{|Z|^2})$.

If $A$ is a self-adjoint operator in Krein space, then similarity of $A$ to a self-adjoint operator in a Hilbert space
admits an equivalent characterization.
Indeed, a characteristic property of a Krein space $(\mathfrak{H}, [\cdot,\cdot]_J)$ is the possibility of its decomposition onto the
direct sum of maximal uniformly
positive $\mathfrak{L}_+$ and maximal uniformly
negative $\mathfrak{L}_-$ subspaces, which are orthogonal with respect to the indefinite metric $[\cdot,\cdot]_J$:
\begin{equation}\label{d2}
\mathfrak{H}=\mathfrak{L}_+[\dot{+}]\mathfrak{L}_-
\end{equation}
(here $[\dot{+}]$ means the orthogonality with respect to the indefinite metric $[\cdot,\cdot]_J$)

The pair of subspaces $\mathfrak{L}_\pm$ in the decomposition (\ref{d2}) is not determined uniquely.

Let $A$ be an operator in $\mathfrak{H}$. We say that the decomposition (\ref{d2})
\emph{is invariant with respect to $A$} if
$$
\mathcal{D}(A)=\mathcal{D}_+[\dot{+}]\mathcal{D}_-, \qquad \mathcal{D}_\pm=\mathcal{D}(A)\cap\mathfrak{L}_\pm
$$
and
$A=A_+[\dot{+}]A_-$, where the operators $A_\pm=A\upharpoonright_{\mathcal{D}_\pm}$ acts in the subspaces $\mathfrak{L}_\pm$, respectively.

\begin{proposition}[\cite{AK1}]\label{sese1}
A pseudo-Hermitian operator $A$ is similar to a self-adjoint operator if and only if there exists
decomposition (\ref{d2}) of the Krein space\footnote{see Sec. 3 for the definition of the Krein space $(\mathfrak{H}, [\cdot,\cdot]_{J{|\eta|}})$.} $(\mathfrak{H}, [\cdot,\cdot]_{J{|\eta|}})$ which is invariant with respect to $A$.
\end{proposition}

The decomposition (\ref{d2}) can be easily characterized with the use of the following operator ${\mathcal C}$:
\begin{equation}\label{bebe5}
{\mathcal C}f={\mathcal C}(f_++f_-)=f_+-f_-, \qquad f=f_++f_-, \quad f_\pm\in\mathfrak{L}_\pm
\end{equation}
(since $\mathfrak{L}_+=(I+{\mathcal C})\mathfrak{H}$ and $\mathfrak{L}_-=(I-{\mathcal C})\mathfrak{H}$).
Therefore, \emph{the invariance of a given decomposition (\ref{d2}) with respect to a linear operator $A$ is equivalent to the relation}
$A{\mathcal C}={\mathcal C}A$.

Assume additionally that $A$ is self-adjoint in a Krein space $(\mathfrak{H}, [\cdot,\cdot]_J)$, then the operators $A_{\pm}$ in the
decomposition $A=A_+[\dot{+}]A_-$ are self-adjoint in the Hilbert spaces $\mathfrak{L}_\pm$ endowed with the inner products ${\pm}[\cdot,\cdot]_J$, respectively.
Therefore, $A$ is self-adjoint in the Hilbert space $\mathfrak{H}$ with the inner product
$$
(f,g)_1=[f_+, g_+]_J-[f_-, g_-]_J, \quad f=f_++f_-, \
g=g_++g_-, \ f_\pm\in\mathfrak{L}_\pm, \ g_\pm\in\mathfrak{L}_\pm.
$$
Taking the definition of $\mathcal{C}$ into account, we get $(\cdot,\cdot)_1=[{\mathcal C}\cdot,\cdot]_J$.
Moreover, it is known (see, for example, \cite{GKS}) that every operator ${\mathcal C}$ defined by (\ref{bebe5}) has the form
${\mathcal C}=Je^Q$, where $Q$ is a \emph{bounded self-adjoint operator in $\mathfrak{H}$ which anticommutes with $J$}:
$QJ=-JQ$. Therefore,
$$
(\cdot,\cdot)_1=[{\mathcal C}\cdot,\cdot]_J=(JJe^Q\cdot,\cdot)=(e^Q\cdot,\cdot)
$$
and, finally we conclude that $A$ is self-adjoint in the Hilbert space $(\mathfrak{H}, (e^Q\cdot,\cdot))$.

\begin{proposition}\label{be77}
A pseudo-Hermitian operator $A$ is similar to a self-adjoint operator if and only there exists an operator
${\mathcal C}=Je^Q$ such that $J=\eta|\eta|^{-1}$, the operator $Q$ satisfies the relations
\begin{equation}\label{bebe7}
Q^*|\eta|=|\eta|Q, \qquad -Q^*\eta={\eta}Q
\end{equation}
and $A{\mathcal C}={\mathcal C}A$. In that case, the operator $A$ turns out to be self-adjoint in the Hilbert space
$\mathfrak{H}$ endowed with inner product $(|\eta|e^{Q}\cdot,\cdot)$.
\end{proposition}
\emph{Proof.}
According to Proposition \ref{sese1} the similarity of $A$ to a self-adjoint operator is
equivalent to the existence of decomposition (\ref{d2}) of the Krein space $(\mathfrak{H}, [\cdot,\cdot]_{J{|\eta|}})$ that
is invariant with respect to $A$. This condition is equivalent to the relation $A{\mathcal C}={\mathcal C}A$, where
${\mathcal C}=Je^Q$ corresponds to the mentioned decomposition of $(\mathfrak{H}, [\cdot,\cdot]_{J{|\eta|}})$.
Taking (\ref{rrr2}) into account, we conclude that $J=\eta|\eta|^{-1}$. Then the relation $QJ=-JQ$ and the condition of
self-adjointness of $Q$ with respect to the inner product $(\cdot,\cdot)_{|\eta|}$ take the form
$$
Q\eta|\eta|^{-1}=-\eta|\eta|^{-1}Q,  \qquad  Q^*|\eta|=|\eta|Q
$$
that is equivalent to (\ref{bebe7}).

The operator $A$ is self-adjoint in the Krein space  $(\mathfrak{H}, [\cdot,\cdot]_{J{|\eta|}})$ and it commutes with
operator ${\mathcal C}=Je^Q$. In that case, as was established above, the operator $A$ is self-adjoint with respect
to the inner product $[{\mathcal C}\cdot,\cdot]_{\eta}=(e^Q\cdot,\cdot)_{|\eta|}=(|\eta|e^Q\cdot,\cdot)$.
 The proof is completed. \rule{2mm}{2mm}

For the case, where $A$ cannot be interpreted as self-adjoint operator in Krein space,
the following general integral-resolvent criterion of similarity can be used:
\begin{lemma}[\cite{NA}]\label{bibi}
 A closed densely defined operator $A$ acting in $\mathfrak{H}$
 is similar to a self-adjoint one if and only if the spectrum of $A$
 is real and there exists a constant $M$ such that
 \begin{equation}\label{bebe84}
 \begin{array}{l}
 \mathrm{sup}_{\varepsilon>0}\varepsilon\int_{-\infty}^{\infty}\|(A-zI)^{-1}g\|^2d\xi\leq{M}\|g\|^2, \vspace{2mm} \\
  \mathrm{sup}_{\varepsilon>0}\varepsilon\int_{-\infty}^{\infty}\|(A^*-zI)^{-1}g\|^2d\xi\leq{M}\|g\|^2,
\quad  \forall{g}\in\mathfrak{H},
 \end{array}
 \end{equation}
 where the integrals are taken along the line $z=\xi+i\varepsilon$ ($\varepsilon>0$
 is fixed) of upper half-plane $\mathbb{C}_+$.
\end{lemma}

 In order to apply Lemma \ref{bibi} to Examples II-IV, we
 need an explicit form of the resolvent $(A_{\mathbf{T}}-zI)^{-1}$. Repeating the proof of
 Lemma 2 in \cite{AK2}, we obtain
 \begin{lemma}\label{l1}
 Let $A_{\mathbf{T}}$ be defined by
 (\ref{lesia800}), (\ref{sas70}) and let $A_0=-d^2/dx^2$, \
 $\mathcal{D}(A_0)=W_2^2(\mathbb{R})$ be the free Schr\"{o}dinger operator in $L_2(\mathbb{R})$.
 Then, for all ${g_{\pm}}\in{L_2(\mathbb{R}_{\pm})}$ and
 for all $z=\tau^2$ from the resolvent set of $A_{\mathbf{T}}$,
 \begin{equation}\label{bbb15}
[(A_{\mathbf{T}}-zI)^{-1}-(A_0-zI)^{-1}]g_{\pm}=c_{1\pm}(\tau)h_{{1\tau}}+c_{2\pm}(\tau)h_{{2\tau}},
\end{equation}
where $h_{{j\tau}}(x)$ are defined by (\ref{sas2}) and
$$
c_{1\pm}(\tau)=\frac{iF_{\pm}(\tau)}{\tau}\left(-1+\frac{2d\tau^2-2i\tau(2\pm{b})}{p(\tau)}\right),
$$
$$
c_{2\pm}(\tau)=\pm\frac{iF_{\pm}(\tau)}{\tau}\left(-1+\frac{-2i\tau(2\mp{c})+2a}{p(\tau)}\right)
$$
where $F_{\pm}(\tau)=\frac{1}{2}\int_{\mathbb{R}}e^{\pm{i}\tau{s}}g_{\pm}(s)ds$
and $p_{\mathbf{T}}(\tau)=2d\tau^2+i(\mathbf{{det} \ T}-4)\tau+2a$.
\end{lemma}

It is known that the resolvent of an arbitrary self-adjoint operator $H$ satisfies the inequality $\|(H-zI)^{-1}\|\leq\frac{1}{|\textsf{Im}\ z|}$
for all $z\in\mathbb{C}\setminus\mathbb{R}$. If $A$ is similar to a self-adjoint operator $H$ (i.e., (\ref{bbbrrr1}) holds), then the inequality above takes the form
\begin{equation}\label{bbb6}
\|(A-zI)^{-1}\|\leq\frac{C}{|\textsf{Im}\ z|}, \qquad C=\|Z^{-1}\|\|Z\|, \quad z\in\mathbb{C}\setminus\mathbb{R}.
\end{equation}
\begin{lemma}\label{b12}
If an operator $A_{\mathbf{T}}$  is similar to a self-adjoint operator in $L_2(\mathbb{R})$, then
the functions
\begin{equation}\label{bbb14}
\Phi_{\pm}(\tau)=\frac{(\textsf{Re}\ \tau)^2}{|\tau|^2}\cdot\frac{|2d\tau^2+i\tau(\mathbf{{det} \ T}\mp2c)|^2+|i\tau(\mathbf{{det} \ T}\pm2b)+2a|^2} {|p_{\mathbf{T}}(\tau)|^2}
\end{equation}
are uniformly bounded on $\mathbb{C}_{++}=\{\tau\in\mathbb{C}_+ : \textsf{Re} \ \tau>0\}$ (i.e.,
there exists $K>0$ such that $\Phi_{\pm}(\tau)<K$ for all $\tau\in\mathbb{C}_{++}$).
\end{lemma}
\emph{Proof.} Let $A_{\mathbf{T}}$ be similar to self-adjoint. Since $A_0$ is self-adjoint, the inequalities  (\ref{bbb6})
hold for $A_{\mathbf{T}}$ and for $A_0$. Therefore,
 for all $g\in{L_2(\mathbb{R})}$ and $z=\tau^2\in{\mathbb{C}_+}$,
 \begin{equation}\label{sas40}
 \|[(A_{\mathbf{T}}-zI)^{-1}-(A_0-zI)^{-1}]g\|^2\leq\frac{M}{(\textsf{Im}\ z)^2}\|g\|^2,
 \end{equation}
 where $M$ is a constant independent of $g$ and $z$.
 In particular, the inequality (\ref{sas40}) holds if we put
 $g=g_+$ or $g=g_-$, where
 $$
 g_+(x)=\left\{\begin{array}{cc}
 e^{-i\overline{\tau}{x}}, & x>0;  \\
 0, & x<0
 \end{array}\right.    \hspace{10mm}
 g_-(x)=\left\{\begin{array}{cc}
 0 , & x>0  \\
 e^{i\overline{\tau}{x}}, & x<0
 \end{array}\right. \qquad \tau\in\mathbb{C}_{++}.
 $$
 In these cases, using (\ref{bbb15}) and taking into account that:
 the functions $h_{j\tau}$ in (\ref{bbb15}) are orthogonal in $L_2(\mathbb{R})$,
 \begin{equation}\label{bebe36}
 \|g_{\pm}\|^2=\frac{1}{2(\textsf{Im}\ \tau)}, \qquad
 \|h_{j\tau}\|^2=\frac{1}{\textsf{Im}\ \tau},
 \qquad  F_{\pm}(\tau)=\frac{1}{4(\textsf{Im}\ \tau)},
 \end{equation}
 and  $(\textsf{Im}\ z)^2=4(\textsf{Im}\ z)^2(\textsf{Re}\ \tau)^2$
we can rewrite (\ref{sas40}) as follows
 $$
 \Phi_{\pm}(\tau)=\frac{(\textsf{Re}\ \tau)^2}{|\tau|^2}M_{\pm}(\tau)\leq{2M}, \quad\forall{\tau}\in\mathbb{C}',
 $$
 where
 $$
 M_{\pm}(\tau)=\left|1-\frac{2d\tau^2-2i\tau(2\pm{b})}{p_{\mathbf{T}}(\tau)}\right|^2+
 \left|1-\frac{-2i\tau(2\mp{c})+2a}{p_{\mathbf{T}}(\tau)}\right|^2
 $$
 Finally, remembering that $p_{\mathbf{T}}(\tau)=2d\tau^2+i{\tau}(\mathbf{{det} \ T}-4)+2a$ we rewrite
 $M_{\pm}(\cdot)$ as
\begin{equation}\label{bebe67}
M_{\pm}(\tau)=\frac{|2d\tau^2+i\tau(\mathbf{{det} \ T}\mp2c)|^2+|i\tau(\mathbf{{det} \ T}\pm2b)+2a|^2} {|p_{\mathbf{T}}(\tau)|^2}
\end{equation}
that gives (\ref{bbb14}). Lemma \ref{b12} is proved. \rule{2mm}{2mm}

The proof of Lemma \ref{bbb6} is close to the part of the proof of Theorem 4 in \cite{AK2}, where the particular case of operators $A_{\mathbf{T}}$ was considered.

\begin{theorem}\label{bbb26}
Let $A\in\{A_{\mathbf{T}}, A_a, A_d\}$ be an operator considered in Examples II-IV.
If the spectrum of $A$ contains the spectral singularity (the exceptional point), then $A$ cannot be similar to a self-adjoint
operator.
\end{theorem}
\emph{Proof.} Assume that $A=A_{\mathbf{T}}$ is a $\mathcal{PT}$-symmetric operator from Example II.
It follows from the proof of Lemma \ref{kkk1} that $A_{\mathbf{T}}$ has a nonzero spectral singularity if
the positive number $\tau=\sqrt{-\frac{a}{d}}$ is the root of $p_{\mathbf{T}}(\tau)$; and $A_{\mathbf{T}}$ has
an exceptional point if the imaginary number $\tau=i\sqrt{-\frac{a}{d}}\in\mathbb{C}_+$ is the root of $p_{\mathbf{T}}(\tau)$ with
multiplicity $2$.

Let us suppose that $A_{\mathbf{T}}$ is similar to self-adjoint. Then, by virtue of Lemma \ref{b12},
the functions $\Phi_{\pm}(\cdot)$ have to be uniformly bounded on $\mathbb{C}_{++}$.
This is impossible since $\Phi_{\pm}(\tau)$ tend to infinity in neighborhood of
$\tau$.

Consider now the case of spectral singularity at point $0$. Then, in view of relations (iv) of Lemma \ref{kkk1},
$$
\Phi_{\pm}(\tau)=\frac{(\textsf{Re}\ \tau)^2}{|\tau|^2}\cdot\frac{|d\tau+i(2\mp{c})|^2+|i(2\pm{b})|^2}{d^2|\tau|^2}
$$
Here $|bc|\not=0$  because $4=\mathbf{{det} \ T}=-bc$. Hence, at least one of functions
$\Phi_{\pm}(\cdot)$ tends to infinity when $\tau\to{0}$.

Finally, if $A_{\mathbf{T}}$ has spectral singularity at $\infty$, then relations (v) of Lemma \ref{kkk1} hold and
$$
\Phi_{\pm}(\tau)=\frac{(\textsf{Re}\ \tau)^2}{|\tau|^2}\cdot\frac{|i\tau(2\mp{c})|^2+|i\tau(2\pm{b})+a|^2} {|a|^2}
$$
It follows from relations (v) that $4=\mathbf{{det} \ T}=-bc$. Hence, $|bc|\not=0$ and at least one of functions
$\Phi_{\pm}(\cdot)$ tends to infinity when $\tau\to{\infty}$.

Summing the cases above we conclude that $A_{\mathbf{T}}$  cannot be similar to a self-adjoint operator.

The cases $A=A_a$ and $A=A_d$ can be considered similarly (it suffices to consider the case of spectral singularity
only). Theorem \ref{bbb26} is proved. \rule{2mm}{2mm}

The functions $M_{\pm}(\tau)$ in (\ref{bebe67}) corresponds to the operator $A_{\mathbf{T}}$
defined by (\ref{lesia800}), (\ref{sas70}). The adjoint operator $A_{\mathbf{T}}^*$ coincides with
$A_{\overline{\mathbf{T}}^t}$. Hence, the following functions:
\begin{equation}\label{bebe67b}
M_{\pm}'(\tau)=\frac{|2\overline{d}\tau^2+i\tau({\mathbf{{det} \ \overline{T}}}\mp2\overline{b})|^2+|i\tau({\mathbf{{det} \ \overline{T}}}\pm2\overline{c})+2\overline{a}|^2} {|p_{\overline{\mathbf{T}}}(\tau)|^2}
\end{equation}
correspond to $A_{\mathbf{T}}^*$.

\begin{theorem}\label{be46}
Let $A_{\mathbf{T}}$ be an operator defined by (\ref{lesia800}), (\ref{sas70}) with real spectrum.  If the functions $M_{\pm}(\tau), M_{\pm}'(\tau)$ are uniformly bounded in
$\mathbb{C}_{++}=\{\tau\in\mathbb{C}_+ : \textsf{Re} \ \tau>0\}$, then $A_{\mathbf{T}}$
is similar to self-adjoint.
\end{theorem}
\emph{Proof.} The operator $A_0$ satisfies relations (\ref{bebe84}) as a self-adjoint operator. Hence,
the inequalities
 \begin{equation}\label{bebe84b}
 \begin{array}{l}
 \mathrm{sup}_{\varepsilon>0}\varepsilon\int_{-\infty}^{\infty}\|[(A_{\mathbf{T}}-zI)^{-1}-(A_0-zI)^{-1}]g\|^2d\xi\leq{M}\|g\|^2, \vspace{3mm} \\
  \mathrm{sup}_{\varepsilon>0}\varepsilon\int_{-\infty}^{\infty}\|[(A_{\mathbf{T}}^*-zI)^{-1}-(A_0-zI)^{-1}]g\|^2d\xi\leq{M}\|g\|^2,
\quad  \forall{g}\in{L_2(\mathbb{R})},
 \end{array}
 \end{equation}
 are necessarily and sufficient condition for the similarity of $A_{\mathbf{T}}$ to a self-adjoint operator.

Let $g=g_+$ be an arbitrary function from $L_2(\mathbb{R}_+)$.
 Using Lemma \ref{l1} and the relation $\|h_{j\tau}(x)\|^2=\frac{1}{\textsf{Im}\ \tau}$
(see (\ref{bebe36})), we get
 \begin{equation}\label{sas66}
 \|[(A_{\mathbf{T}}-zI)^{-1}-(A_0-zI)^{-1}]g_+\|^2=\frac{|F_+(\tau)|^2}{|\tau|^2(\textsf{Im}\ \tau)}M_+(\tau),
 \end{equation}
 where $F_+(\tau)$ is the Fourier transform of $g_+$ and $z=\tau^2$ $(\tau\in\mathbb{C}_{++})$.

 Since $M_+(\tau)$ is uniformly bounded on $\mathbb{C}_{++}$, there is a constant $K_1>0$ such that
 $|M_+(\tau)|\leq{K_1}$.
 Then
 \begin{equation}\label{des1}
\varepsilon\int_{-\infty}^{\infty}\|[(A_{\mathbf{T}}-zI)^{-1}-(A_0-zI)^{-1}]g_+\|^2d\xi{\leq}
K_1\int_{-\infty}^{\infty}\frac{\varepsilon|F_+(\tau)|^2}{|\tau|^2(\textsf{Im}\ \tau)}d\xi.
\end{equation}

Let us consider an auxiliary self-adjoint operator $\tilde{A}_{\mathbf{T}}$ with $b=c=0$, $a=1$, and $d=4$. Then $\mathbf{{det} \ T}=4$ and
$$
\tilde{M}_{+}(\tau)=\frac{|\gamma^2+i\gamma|^2+|i\gamma+1|^2}{|\gamma^2+1|^2}=\frac{|\gamma|^2}{|\gamma-i|^2}+\frac{1}{|\gamma+i|^2}, \quad \gamma=2\tau\in\mathbb{C}_{++}.
$$
The obtained expression leads to the conclusion that $\tilde{M}_{+}(\tau)\geq\frac{1}{4}$ for all $\tau\in\mathbb{C}_{++}$.
Taking this inequality into account and using (\ref{bebe84b}) and (\ref{sas66})
for the pair of self-adjoint operators  $\tilde{A}_{\mathbf{T}}$, $A_0$, we obtain
\begin{eqnarray*}
\frac{1}{4}\int_{-\infty}^{\infty}\frac{\varepsilon|F_+(\tau)|^2}{|\tau|^2(\textsf{Im}\ \tau)}d\xi\leq{\int_{-\infty}^{\infty}\frac{\varepsilon|F_+(\tau)|^2}{|\tau|^2(\textsf{Im}\ \tau)}\tilde{M}_+(\tau)d\xi}= & & \\
\varepsilon\int_{-\infty}^{\infty}\|[(\tilde{A}_{\mathbf{T}}-zI)^{-1}-(A_0-zI)^{-1}]g_+\|^2d\xi<M\|g_+\|^2, & &
\end{eqnarray*}
where $M$ is a constant independent of $\varepsilon>0$ and $g_+$.

Combining the obtained evaluation with (\ref{des1}), we obtain
$$
\varepsilon\int_{-\infty}^{\infty}\|[(A_{\mathbf{T}}-zI)^{-1}-(A_0-zI)^{-1}]g_+\|^2d\xi<4K_1M\|g_+\|^2,
$$
where $4K_1M$ does not depend on $\varepsilon>0$ and $g_+$.

Considering similarly the case  $g=g_-(x)\in{L_2(\mathbb{R}_-)}$ (here the uniformly boundedness of $M_{-}(\tau)$ has to be used)
and, consequently,  the case of operator
$A_{\mathbf{T}}^*$,  we arrive at the conclusion that (\ref{bebe84b}) hold for all
functions from $L_2(\mathbb{R})$. Hence, $A$ is similar
to a self-adjoint operator. Theorem \ref{be46} is proved.
\rule{2mm}{2mm}

\begin{corollary}\label{be46b}
Let $A_{\mathbf{T}}$ satisfy conditions of Theorem \ref{be46}.
If, in addition, $A_{\mathbf{T}}$ can be interpreted as a self-adjoint operator in a Krein space, then
the property of $M_{\pm}(\tau)$ to be uniformly
bounded in $\mathbb{C}_{++}$ implies the
similarity of $A_{\mathbf{T}}$ to a self-adjoint operator.
\end{corollary}
\emph{Proof.} If $A_{\mathbf{T}}$ can be interpreted as self-adjoint in a Krein space, then, for a certain choice of
fundamental symmetry $J$, the equality (\ref{bbb23}) holds for $A_{\mathbf{T}}$ and $A_{\mathbf{T}}^*$. In that case, the first and the second inequalities in (\ref{bebe84}) are equivalent.
Obviously, the same remains true for the inequalities (\ref{bebe84b}). Thus, for the similarity of $A_{\mathbf{T}}$
to a self-adjoint operator it suffices to establish the first inequality in (\ref{bebe84b}). The latter is
ensured by uniformly boundedness property of $M_{\pm}(\tau)$ in $\mathbb{C}_{++}$
(see the proof of Theorem  \ref{be46}). \rule{2mm}{2mm}
\vspace{3mm}

{\bf Example II contd.}
\begin{corollary}\label{bbb27b}
Let $A_{\mathbf{T}}$  be a $\mathcal{PT}$-symmetric operator considered in Example II.
If one of the following conditions is satisfied, then $A_{\mathbf{T}}$ is similar to a self-adjoint operator:
\begin{enumerate}
  \item[(i)] \quad $D=(4-\mathbf{{det} \ T})^2+16ad<0$, \quad $(4-\mathbf{{det} \ T})d<0$;
  \item[(ii)] \quad $D=0$, \quad $(4-\mathbf{{det} \ T})d<0$;
\end{enumerate}
\end{corollary}
\emph{Proof.}
Every condition (i), (ii) guarantees that $d\not=0$ and the roots $\tau_{1,2}$ of the polynomial $p_{\mathbf{T}}(\tau)$ (see (\ref{bbb1})) belong to $\mathbb{C}_{-}$. Then the functions $M_{\pm}(\tau)$ (see (\ref{bebe67})) are uniformly bounded in $\mathbb{C}_{++}$.
By Proposition \ref{hhh1}, $A_{\mathbf{T}}$ can be realized as self-adjoint in a Krein space.
Hence, we can apply Corollary \ref{be46b} that completes the proof. \rule{2mm}{2mm}

The conditions of Corollary \ref{bbb27b} ensure the uniformly boundedness of  $M_{\pm}(\tau)$.
This property is \emph{sufficient} for the similarity of $A_{\mathbf{T}}$ to a self-adjoint operator.
If $D>0$ the corresponding roots $\tau_{1,2}$ in (\ref{bbb1}) lie on imaginary axes $i\mathbb{R}$ and may happen that at least one of them
(let, for definiteness, $\tau_1$) belongs to $\mathbb{C}_+$. In that case the functions $M_{\pm}(\tau)$ may tend to infinity as $\tau\to\tau_1$. However, as we show below, the corresponding operator $A_{\mathbf{T}}$ remains similar to a self-adjoint operator.

\begin{theorem}\label{fff1}
Let $A_{\mathbf{T}}$  be a $\mathcal{PT}$-symmetric operator considered in Example II and let
$D=(4-\mathbf{{det} \ T})^2+16ad>0$. Then $A_{\mathbf{T}}$
is similar to a self-adjoint operator.
\end{theorem}
\emph{Proof.}  By virtue of Proposition \ref{hhh1}, $A_{\mathbf{T}}$ is
self-adjoint in the Krein space $(L_2(\mathbb{R}), [\cdot,\cdot]_{\mathcal{P}_\phi})$.
Using Proposition \ref{be77} with $\eta=J=\mathcal{P}_\phi$ we conclude that the similarity of $A_{\mathbf{T}}$ to a
self-adjoint operator in a Hilbert space is equivalent to the existence of an operator ${\mathcal C}=\mathcal{P}_\phi{e}^{Q}$
which satisfies the following conditions:
\begin{equation}\label{bebe38}
Q^*=Q, \qquad \mathcal{P}_{\phi}Q=-Q\mathcal{P}_{\phi}, \qquad A{\mathcal C}={\mathcal C}A.
\end{equation}

Let $Q=\chi{i}\mathcal{R}\mathcal{P}_{\phi}$, where $\chi\in\mathbb{R}$ and $\mathcal{R}$ be defined by (\ref{bbb34}).
The fundamental symmetry $\mathcal{R}$ anti-commutes with $\mathcal{P}$ and hence, $\mathcal{R}$ anti-commutes with
the fundamental symmetry $\mathcal{P}_\phi=\mathcal{P}e^{i\phi\mathcal{R}}$. This means that
$Q$ satisfies the first two conditions of (\ref{bebe38}).
The third condition is equivalent to the relation
\begin{equation}\label{bebe39}
A_{\mathbf{T}}^*e^Q=e^QA_{\mathbf{T}}
\end{equation}
since $A_{\mathbf{T}}\mathcal{P}_\phi=\mathcal{P}_{\phi}A_{\mathbf{T}}^*$ and ${\mathcal C}=\mathcal{P}_\phi{e}^{Q}$.

The operator ${i}\mathcal{R}\mathcal{P}_{\phi}$ is a fundamental symmetry in $L_2(\mathbb{R})$ because $\mathcal{R}$ anti-commutes with
$\mathcal{P}_\phi$. This property allows us to rewrite $e^Q$ as
\begin{equation}\label{bbb28}
e^Q=e^{\chi{i}\mathcal{R}\mathcal{P}_{\phi}}=[\cosh{\chi}]I+[\sinh{\chi}]{i}\mathcal{R}\mathcal{P}_{\phi}.
\end{equation}
The obtained expression shows that $e^Q$ commutes with the symmetric operator
$A_{\mathrm{sym}}$ defined by (\ref{tato1}) and commutes with the adjoint operator $A_{\mathrm{sym}}^*$.
Hence, (\ref{bebe39}) holds if
$e^Q : \mathcal{D}(A_{\mathbf{T}})\to\mathcal{D}(A_{\mathbf{T}}^*)$.
The latter relation
is equivalent to the following implication
\begin{equation}\label{lesia701}
\mbox{if} \ \hspace{5mm} \mathbf{T}\Gamma_0f=\Gamma_1f,
\hspace{5mm} \ \mbox{then} \ \hspace{5mm}
\overline{\mathbf{T}}^{\mathrm{t}}\Gamma_0{e^{\chi{i}\mathcal{R}\mathcal{P}_{\phi}}}f=\Gamma_1{e^{\chi{i}\mathcal{R}\mathcal{P}_{\phi}}}f,
\end{equation}
where $\Gamma_j$ are boundary operators from Lemma \ref{be1} and $f$ is an arbitrary element of $\mathcal{D}(A_{\mathbf{T}})$.

Thus \emph{if (\ref{lesia701}) holds for a certain $\chi\in\mathbb{R}$, then $A_{\mathbf{T}}$ is similar to a self-adjoint operator in a Hilbert space.}

Denote
$$
\sigma_1=\left(\begin{array}{cc} 0  & 1 \\
1 & 0
\end{array}\right), \quad \sigma_2=\left(\begin{array}{cc} 0  & -i \\
i & 0
\end{array}\right), \quad \sigma_3=\left(\begin{array}{cc} 1  & 0 \\
0 & -1
\end{array}\right).
$$

Using (\ref{sas70}), (\ref{bbb11}), (\ref{bbb34}), and (\ref{bbb28}) it is easy to verify that
$$
\Gamma_0{e^Q}f=\cosh{\chi}\Gamma_0f+\sinh{\chi}[-\frac{i}{2}\cos\phi\sigma_1\Gamma_1f+\sin\phi\sigma_3\Gamma_0f],
$$
and
$$
\Gamma_1{e^Q}f=\cosh{\chi}\Gamma_1f+\sinh{\chi}[{2i}\cos\phi\sigma_1\Gamma_0f+\sin\phi\sigma_3\Gamma_1f]
$$
for all $f\in\mathcal{D}(A_{\mathrm{sym}}^*)={W_2^2(\mathbb{R}\backslash\{0\})}$.

Substituting these expressions into (\ref{lesia701}),
we obtain  that (\ref{lesia701})
is equivalent to the matrix equality:
\begin{equation}\label{bebe67c}
\cosh{\chi}(\mathbf{T}-\overline{\mathbf{T}}^{\mathrm{t}})
=-\sinh{\chi}(\frac{i}{2}\cos\phi[\overline{\mathbf{T}}^{\mathrm{t}}\sigma_1\mathbf{T}+4\sigma_1]+\sin\phi[\sigma_3\mathbf{T}-\overline{\mathbf{T}}^{\mathrm{t}}\sigma_3]).
\end{equation}

Since $A_{\mathbf{T}}$ is $\mathcal{PT}$-symmetric, the entries of $\mathbf{T}$ satisfy (\ref{bbb35}) and we
can set $b=ix$, $c=iy$, where $x,y$ are arbitrary real numbers.

Simple calculations give:
$$
\mathbf{T}-\overline{\mathbf{T}}^{\mathrm{t}}=i(x+y)\sigma_1, \quad \overline{\mathbf{T}}^{\mathrm{t}}\sigma_1\mathbf{T}+4\sigma_1=(\mathbf{{det} \ T}+4)\sigma_1, \quad \sigma_3\mathbf{T}-\overline{\mathbf{T}}^{\mathrm{t}}\sigma_3=i(x-y)\sigma_1.
$$
Hence, the matrix relation (\ref{bebe67c}) can be reduced to the equality
$$
x+y=-\frac{\sinh{\chi}}{\cosh{\chi}}\left[\frac{1}{2}(\mathbf{{det} \ T}+4)\cos\phi+(x-y)\sin\phi\right],
$$
which, obviously, has a solution $\chi$ if and only if
\begin{equation}\label{bebe79}
\left[\frac{1}{2}(\mathbf{{det} \ T}+4)\cos\phi+(x-y)\sin\phi\right]^2>(x+y)^2.
\end{equation}

Using the identity $2(x-y)\cos\phi=(\mathbf{{det} \ T}+4)\sin\phi$, which follows
directly from (\ref{rere1}), and making elementary transformations, we reduce
(\ref{bebe79}) to the inequality
\begin{equation}\label{bebe80}
(\mathbf{{det} \ T}+4)^2-16xy>0.
\end{equation}
To complete the proof it is sufficient to observe that the inequality (\ref{bebe80}) coincides with the
condition $D>0$ (since (\ref{bbb17}) and $b=ix, c=iy$). Theorem \ref{fff1} is proved.
\rule{2mm}{2mm}
\smallskip

Summing the results above we obtain the following relationship between properties of
$A_{\mathbf{T}}$ and the parameters  $D=(4-\mathbf{{det} \ T})^2+16ad$, \ $K=(4-\mathbf{{det} \ T})d$.
\begin{center}
\begin{tabular}{|c|c|c|c|}  \hline
 {} & $K>0$ & $K=0$ & $K<0$   \\ \hline
 $D>0$ & similarity & similarity & similarity   \\ \hline
 $D=0$ & exceptional point & $\begin{array}{c}  \mbox{spectral singularity at} \ 0 \\
 \mbox{spectral singularity at} \ \infty \\
 \sigma(A_{\mathbf{T}})=\mathbb{C} \end{array}$  & similarity   \\ \hline
 $D<0$ & pair of complex eigenvalues & nonzero spectral singularity & similarity   \\ \hline
\end{tabular}
\end{center}

\smallskip

{\bf Examples III. VI contd.}
\begin{corollary}\label{bbb27}
Let $A_a$ ($A_d$) be an operator considered in Example III (IV).
If $\textsf{Re}\ a>0$ \ ($\textsf{Re}\ d<0$), then $A_a$  ($A_d$) is similar to a self-adjoint operator.
\end{corollary}
\emph{Proof.} Let $\textsf{Re}\ a>0$. Then the spectrum of $A_a$ is real. The adjoint $A_a^*$ coincides with $A_{\overline{a}}$.
The functions $M_{\pm}(\tau), M_{\pm}'(\tau)$ have the form
$$
M_{+}(\tau)=M_{-}(\tau)=\frac{|a|^2}{|-2i\tau+a|^2}, \quad  M_{+}'(\tau)=M_{-}'(\tau)=\frac{|\overline{a}|^2}{|-2i\tau+\overline{a}|^2}.
$$

If $\textsf{Re}\ a>0$, then the roots  $\displaystyle{\tau_{1}=-\frac{ia}{2}}, \ \displaystyle{\tau_{2}=-\frac{i\overline{a}}{2}}$ of the denominators belong to $\mathbb{C}_-$. In these cases, $M_{\pm}(\tau), M_{\pm}'(\tau)$ are uniformly bounded in $\mathbb{C}_{++}$.
 By Theorem \ref{be46},  the operator $A_a$ is similar to self-adjoint.

The operators $A_d$ have real spectrum when $\textsf{Re}\ d<0$ and $A^*_d=A_{\overline{d}}$. The functions
$$
M_{+}(\tau)=M_{-}(\tau)=\frac{|d\tau|^2}{|d\tau-2i|^2}, \quad  M_{+}'(\tau)=M_{-}'(\tau)=\frac{|\overline{d}\tau|^2}{|\overline{d}\tau-2i|^2}.
$$
are uniformly bounded in $\mathbb{C}_{++}$. Using again Theorem \ref{be46} we complete the proof.
 \rule{2mm}{2mm}

\smallskip

The following picture illustrates the change of properties of $A_a$ (complex eigenvalue $\to$ spectral singularity $\to$ similarity to a self-adjoint operator):

\smallskip

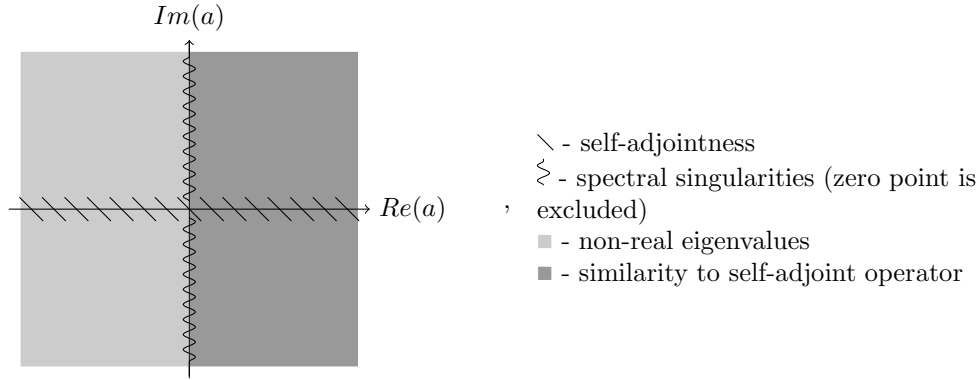
\begin{center}
\begin{tikzpicture}[scale=1.5]
  \filldraw[fill=black!20, draw=white] (-1.5cm, -1.4cm) rectangle (0, 1.4cm);
  \filldraw[fill=black!40, draw=white] (0, -1.4cm) rectangle (1.5cm, 1.4cm);

  \draw [->] (-1.6cm, 0) -- (1.6cm, 0) node[right] {$Re(a)$};
  \draw [->] (0, -1.5cm) -- (0, 1.5cm) node[above] {$Im(a)$};

  \foreach \x in {-1.4cm,-1.2cm,...,1.4cm}
    \draw (\x-3pt,3pt) -- (\x+3pt,-3pt);

  \draw [decorate,decoration={snake,amplitude=0.8mm,segment length=2mm,post length=1mm}]
        (0,2pt) -- (0,1.48cm);
  \draw [decorate,decoration={snake,amplitude=0.8mm,segment length=2mm,post length=1mm}]
        (0,-2pt) -- (0,-1.48cm);

%
%
%

\draw[xshift=3cm]
  node[left, text width=0.2cm]
    {
     ,
    }
  node[right,text width=6cm]
    {
      \begin{tikzpicture}
        \draw (-3pt,3pt) -- (3pt,-3pt);
      \end{tikzpicture} - self-adjointness\\

      \begin{tikzpicture}
        \draw [decorate,decoration={snake,amplitude=0.8mm,segment length=2mm,post length=0}]
        (0,0) -- (0,10pt);
      \end{tikzpicture} - spectral singularities (zero point is excluded)\\

      \begin{tikzpicture}
        \filldraw[fill=black!20, draw=white] (0.15cm,-0.15cm) rectangle (0.34cm, -0.34cm);
      \end{tikzpicture} - non-real eigenvalues\\

      \begin{tikzpicture}
        \filldraw[fill=black!40, draw=white] (0.15cm,-0.15cm) rectangle (0.34cm, -0.34cm);
      \end{tikzpicture} - similarity to self-adjoint operator\\
    };
\end{tikzpicture}
\end{center}


\begin{thebibliography}{AL1}
\bibitem{ADKK}
{S. Albeverio, M. Dudkin, A. Konstantinov and V. Koshmanenko},
\textit{On the point spectrum of $\mathcal{H}_{-2}$-singular perturbations},
{Mathematische Nachrichten}
\textbf{12} 
{(2006)},
{no.~280}, 
{20--27}. 

\bibitem{AKK}
{S. Albeverio, A. Konstantinov and V. Koshmanenko},
\textit{The Aronszajn–Donoghue Theory for Rank One Perturbations of the ${\mathcal H}_{-2}$-class},
{Integral Equations and Operator Theory}
\textbf{50}
{(2004)},
{no.~1},
{1--8}.

\bibitem{AL1}
{S. Albeverio and P. Kurasov},
\textit{Singular Perturbations of Differential Operators and Solvable Schr\"{o}dinger Type Operators},
{London Math. Soc. Lecture Note Ser.}
\textbf{271}
{Cambridge: Cambridge University Press}
{(2000)}

\bibitem{AK1}
{S. Albeverio and S. Kuzhel},
\textit{Pseudo-Hermiticity and theory of singular perturbations},
{Lett. Math. Phys.}
\textbf{67}
{(2004)},
{no.~3},
{223--238}.

\bibitem{AK2}
{S. Albeverio and S. Kuzhel},
\textit{One-dimensional Schr\"{o}dinger operators with ${\mathcal P}$-symmetric zero-range potentials},
{J. Phys. A.}
\textbf{38}
{(2005)},
{4975--4988}.

\bibitem{AZ}
{T.\ Ya. Azizov and I.S. Iokhvidov},
\textit{Linear Operators in Spaces with an Indefinite Metric},
{John Wiley \& Sons},
{Chichester},
{(1989)}.

\bibitem{B6}
{C. M. Bender},
\textit{Introduction to PT-Symmetric Quantum Theory},
{Contemp.Phys.}
\textbf{46}
{(2005)},
{277--292}.

\bibitem{B4}
{C. M. Bender},
\textit{Making sense of non-Hermitian Hamiltonians},
{Rep. Prog. Phys.}
\textbf{70}
{(2007)},
{947--1018}.

\bibitem{HK}
{S. Hassi and S. Kuzhel},
\textit {On J-self-adjoint operators with stable C-symmetries},
{Proceedings of the Royal Society of Edinburgh: Section A Mathematics}
\textbf{143}
{(2013)},
{no.~1},
{141--167}.

\bibitem{GKS}
{A. Grod, S. Kuzhel, and V. Sudilovskaja},
\textit{On operators of transsition in Krein spaces},
{Opuscula Mathematica}
\textbf{31}
{(2011)},
{no.~1},
{49--59}.

\bibitem{GR}
{U. G\"unther, I. Rotter, and B. Samsonov},
\textit{Projective Hilbert space structures at exceptional points},
{J. Phys. A.}
\textbf{40}
{(2007)},
{8815--8833}. 

\bibitem{KKQ}
{W. Karwowski, V. Koshmanenko and S. Ota},
\textit{Schrodinger Operator Perturbed by Operators Related to Null Sets},
{Positivity}
\textbf{2}
{(1998)},
{no.~1},
{77--99}.

\bibitem{KO}
{V. Koshmanenko},
\textit{Singular Quadratic Forms in Perturbation Theory},
{Mathematics and Its Applications}
\textbf{474}
{Kluwer Academic Publishers},
{Dordrecht},
{(1999)}

\bibitem{PK}
{S.O. Kuzhel, O.M. Patsiuk},
\textit{Interpretatation of ${\mathcal P}{\mathcal T}$-summetric operators like self-adjoint ones in Krein spaces (Ukranian)},
{Collection of research works of Institute of Mathematics of NAS of Ukraine}
\textbf{8}
{(2011)}
{no.~1},
{111--127}.

\bibitem{MO1}
{A. Mostafazadeh},
\textit{Delta-function potential with a complex coupling},
{J. Phys. A: Math. Gen}
\textbf{39}
{(2006)},
{13495--13506}.

\bibitem{MO}
{A. Mostafazadeh},
\textit{Pseudo-Hermitian Representation of Quantum Mechanics},
{Int. J. Geom. Meth. Mod. Phys.}
\textbf{7}
{(2010)},
{1191-1306}.

\bibitem{M1}
{A. Mostafazadeh},
\textit{Pseudo-Hermiticity versus ${\mathcal P}{\mathcal T}$-symmetry: the necessary condition for the reality of the spectrum of a non-Hermitian Hamiltonian},
{J. Math. Phys.}
\textbf{43}
{(2002)},
{205--214}.

\bibitem{NA}
{S. Naboko},
\textit{Conditions for similarity to unitary and selfadjoint operators (Russian)},
{Func. Anal. and Appl.}
\textbf{18}
{(1984)},
{no.~1},
{16--27}.
\end{thebibliography}
\end{document}